\documentclass[
superscriptaddress,
twocolumn,
amsmath,amssymb,
aps,
prx,
nofootinbib,
]{revtex4-2}

\usepackage{comment}
\usepackage{tabularx}
\usepackage{graphicx}
\usepackage{dcolumn}
\usepackage{bm}
\usepackage{mathtools}
\usepackage{color}
\usepackage{braket}
\usepackage{gensymb}
\usepackage{lineno}
\usepackage{xr}  

\usepackage{hyperref}
\hypersetup{
	colorlinks=true,
	linkcolor=blue,
	citecolor=blue,
	urlcolor=blue
}

\begin{document}

\title{Parity-Doublet Coherence Times in Optically Trapped Polyatomic Molecules}

\author{Paige Robichaud}
\email{paigerobichaud@g.harvard.edu}
\author{Christian Hallas}
\author{Junheng Tao}
\author{Giseok Lee}
\author{Nathaniel B. Vilas}
\author{John M. Doyle}
\affiliation{Department of Physics, Harvard University, Cambridge, MA 02138, USA}
\affiliation{Harvard-MIT Center for Ultracold Atoms, Cambridge, MA 02138, USA}

\date{\today}

\begin{abstract}
Polyatomic molecules provide complex internal structures that are ideal for applications in quantum information science, quantum simulation, and precision searches for physics beyond the Standard Model. A key feature of polyatomic molecules is the presence of parity-doublet states. These structures, which generically arise from the rotational and vibrational degrees of freedom afforded by polyatomic molecules, are a powerful feature to pursue these diverse quantum science applications. Linear triatomic molecules contain $\ell$-type parity doublet states, which are predicted to exhibit robust coherence properties. We optically trap CaOH molecules, prepare them in $\ell$-type parity-doublet states, and realize a bare qubit coherence time of $T_2^* = 0.8(2)$~s. We suppress differential Stark shifts by employing molecular spectroscopy to cancel ambient electric fields, and characterize  parity-dependent trap shifts, which are found to limit the coherence time. The parity-doublet coherence times achieved in this work are a defining milestone for the use of polyatomic molecules in quantum science. 

\end{abstract}

\maketitle

Ultracold molecules are powerful tools for quantum science due to their structural complexity. Large intrinsic dipole moments and rich Hilbert spaces enable novel and robust schemes for quantum simulation~\cite{wall2015quantum, micheli2006toolbox,Gorshkov2011}, quantum information processing~\cite{demille2002quantum,tesch2002quantum,wei2011entanglement,Yu2019}, and precision searches for physics beyond the Standard Model~\cite{kozyryev2017PolyEDM,Norrgard2019,Hutzler2020,kozyryev2021enhanced,demille2024quantum}. Progress towards these goals with direct laser cooling of molecules~\cite{Barry2014, anderegg2017CaFMOT, truppe2017CaF,Collopy2018,zeng2024three} and indirect assembly of molecules~\cite{ni2008high,koppinger2014production,Takekoshi2014,Park2015,Guo2016,Rvachov2017Long,voges2020ultracold,Cairncross2021Assembly} has matured the platform to several realizations of functional optical tweezer arrays of ultracold molecules \cite{Anderegg2019Optical,Cairncross2021Assembly,Zhang2022optical,ruttley2023formation,holland2023bichromatic}, where single quantum state control~\cite{park2017secondscale,Burchesky2021Rotational} and entanglement~\cite{holland2023demand,bao2023dipolar,picard2025entanglement, ruttley2025long} have been demonstrated. The rotational and vibrational energy structure of molecules harbors a selection of strongly interacting and long-lived states that make versatile qubits~\cite{tesch2002quantum,Sawant2019,albert2019robust}. In several species of diatomic molecules, long coherence times have been achieved in hyperfine storage qubits~\cite{park2017secondscale,gregory2021robust} and interacting rotational qubits~\cite{Burchesky2021Rotational, park2023extended,Gregory2024}. These results took advantage of a ``magic" trapping light condition, where differential light shifts between qubit states are tuned close to zero with the polarization~\cite{neyenhuis2012anisotropic, seesselberg2018extending, Burchesky2021Rotational, gregory2021robust, park2023extended}, intensity~\cite{Blackmore2018}, or wavelength of the light~\cite{Bause2020tuneout,guan2021magic,Gregory2024}. Enabled by long rotational coherence times, high-fidelity dipole-dipole interactions, including entangling iSWAP gate operations, have been recently demonstrated in diatomic molecules~\cite{holland2023demand,bao2023dipolar,picard2025entanglement,ruttley2025long}.

Polyatomic molecules possess additional, nuanced structural features compared to diatomics. One key advantage of polyatomic molecules is the presence of pairs of long-lived, near-degenerate states of opposite parity called ``parity-doublets.'' These states have been proposed for a number of novel applications in quantum science and precision measurements of fundamental physics. For example, parity-doublet states fully mix with modest applied electric fields, resulting in states with large, saturated lab-frame electric dipole moments along with states having zero dipole moment. This structure enables fast and strong switchable electric dipole interactions for novel quantum computing schemes~\cite{wei2011entanglement, Yu2019} and is a natural platform to explore models of quantum magnetism, including integer spin systems~\cite{Wall2013, Wall2015, wall2015quantum}. The Stark structure also allows for internal co-magnetometer states robust to systematic effects and advantageous to searches for physics beyond the Standard Model (BSM)~\cite{kozyryev2017PolyEDM,Hutzler2020,takahashi2023engineering,takahashi2025engineered}. At zero electric field, dipole-dipole interactions between parity-doublet state molecules can be used to construct iSWAP two-qubit gates~\cite{holland2023demand,bao2023dipolar,picard2025entanglement,ruttley2025long}.

The above applications require robust coherence properties. Parity-doublet states promise long coherence times because they share all quantum numbers—except for parity—and therefore suppress dephasing effects from environmental perturbations. In linear triatomic molecules, parity doublets arise from a degeneracy of the projection of vibrational angular momentum onto the internuclear axis, $\ell$, as shown in Figure $\ref{fig:1}$. These $\ell$-type parity doublets possess, to high order, identical magnetic field sensitivities due to their common electron spin, nuclear spin, and rotational angular momentum. Similarly, differential light shifts in $\ell$-type parity doublets are reduced to only parity-dependent effects. At low electric fields, the Stark sensitivity of parity-doublets is a quadratic differential shift, making them robust to dephasing from fluctuations in the environmental electric field. Motivated by the advantages of parity-doublet states, among other properties, polyatomic molecules have been laser cooled~\cite{kozyryev2016Sisyphus,augenbraun2023review}, trapped~\cite{Zeppenfeld2012Sisyphus,prehn2016,liu2017magnetic,vilas2022magneto,hallas2022optical,Lasner2024MOT,sawaoka2025optical}, loaded into optical tweezer arrays~\cite{VilasOpticalTweezer2024}, and characterized for quantum science applications~\cite{anderegg2023quantum,low2025coherence}.

In this work, we measure the coherence time of parity-doublet states in optically trapped, ultracold polyatomic molecules. Two pairs of fully-stretched parity doublets in CaOH are examined, with each doublet belonging to either the $N=1$ or $N=2$ rotational state in an excited bending vibrational state, $\tilde{X}(010)$, shown in Figure \ref{fig:1}. To mitigate decoherence arising from differential quadratic Stark shifts, we cancel ambient electric fields using Hz-level spectroscopy of the parity-doublet transition. We measure and compare the rotational dependence of the Stark sensitivity in the parity-doublet states. The coherence time of both doublets is found to be limited by light shifts from the optical trap, which are reduced with a magic polarization angle. We observe a bare coherence time of $T_2^* = 0.8(2)~\text{s}$ in the $N=1$ parity-doublet states.

\begin{figure}
    \centering
    \includegraphics[width=\linewidth]{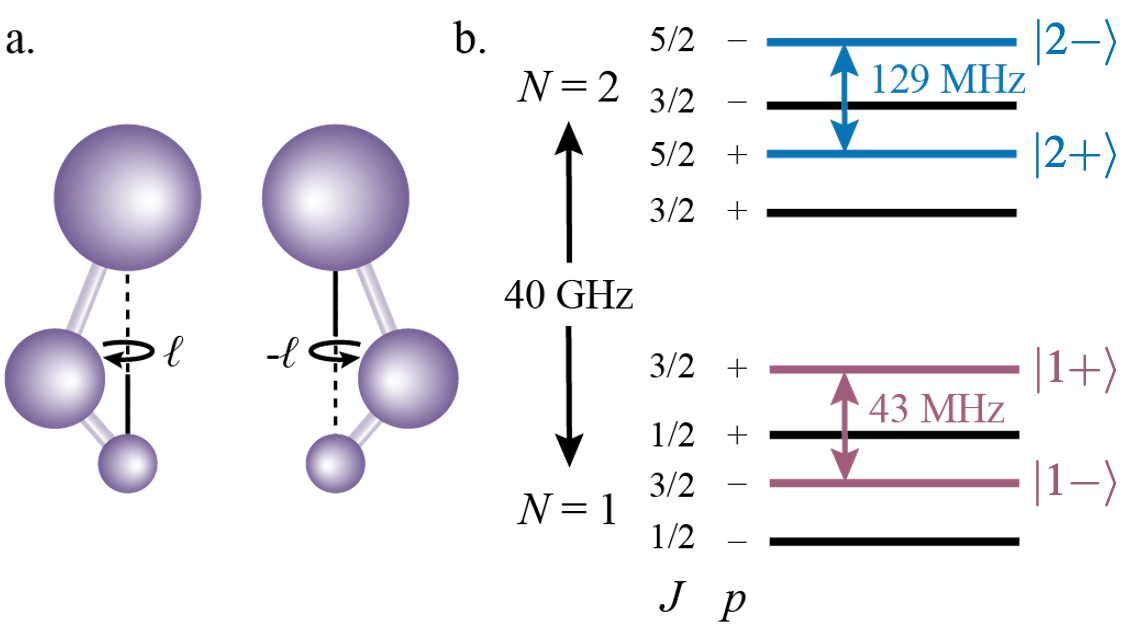}
    \caption{a. A visual depiction of the vibrational bending mode of the molecule. 
    The projection of the vibrational angular momentum onto the internuclear axis is given by $\ell$. b. $N=1$ and $N=2$ rotational state structure in the $\tilde{X}(010)$ vibronic state of CaOH, containing parity doublet states. The degeneracy between parity doublet states is split by Coriolis interactions which separate in-plane and out-of-plane vibrations, where $\ket{\pm} = \frac{1}{\sqrt{2}}\left(\ket{\ell}\pm(-1)^{N-\ell}\ket{-\ell}\right)$. The qubit states examined in this work are fully stretched states labeled $\ket{1+}$ and $\ket{1-}$ in $N=1$ and $\ket{2+}$ and $\ket{2-}$ in $N=2$.}
    \label{fig:1}
\end{figure}

Our experiment starts with loading laser cooled CaOH molecules into a $1064~\text{nm}$ optical dipole trap using single-frequency gray molasses cooling following blue-detuned magneto-optical trapping~\cite{Hallas2024}. The molecules are then optically pumped into the vibrational bending mode and prepared in a single hyperfine parity-doublet state using optical, microwave, and radio frequency (RF) pulses, as described in the Supplemental Material~\cite{Supplemental}. We choose fully-stretched parity-doublet states to maximize their interaction strength. We label them for $N=1$ as $\ket{1\pm} \equiv \ket{N=1, J=3/2, F=2, m_F = 2, p=\pm}$. As the parity doublets in $N=2$ are expected to have lower Stark sensitivity compared with $N=1$, we also examine the coherence of these $N=2$ states, labeled $\ket{2\pm} \equiv \ket{N=2, J=5/2, F=3, m_F = 3, p=\pm}$. Here $N$ is the rotational quantum number, $J$ is the total angular momentum excluding nuclear spin, $F$ is the total angular momentum including nuclear spin, and $p$ is the parity of the state. All states are spectroscopically resolvable with a bias magnetic field, which is ramped from $3~\text{G}$ to higher fields after preparation in the parity-doublet states. Prior to the Ramsey measurement of coherence time, the trap depth is lowered from $U_0\sim640~\text{\textmu K}$ to $U_0\sim40~\text{\textmu K}$, which adiabatically cools the molecules from $\sim$$60~\text{\textmu K}$ to $\sim$$15~\text{\textmu K}$.

We perform a Ramsey sequence to measure the coherence time, $T_2^*$, of the parity-doublet states. The sequence consists of a pair of RF $\pi/2$ pulses separated by a Ramsey hold time, $\tau$. The RF is applied with two in-vacuum coils at $\sim$$43~\text{MHz}$ and $\sim$$129~\text{MHz}$ for the $N=1$ and $N=2$ parity-doublet transitions, respectively (these coils are also used to drive the RF MOT field). We measure the Ramsey fringe contrast in two ways, for technical convenience. For the data shown in Figure \ref{fig:2}, the RF is detuned by 1~kHz and the precession time is $\tau + T$, where the fringe measurement time $T$ is scanned over a period of 1~ms to determine the contrast. In the data shown in Figures \ref{fig:3} and \ref{fig:4}, the RF is resonant and the relative phase of the second Ramsey $\pi/2$ pulse is scanned over a period from $0$ to $2\pi$ to measure the Ramsey fringe. After the Ramsey measurement, the molecular population may occupy the $\ket{+}$ or $\ket{-}$ parity-doublet state in addition to a distribution of rovibrational states as a result of blackbody radiation and radiative decay. To selectively measure only the parity-doublet population, we apply a ``pushout" pulse of resonant light that ejects all detectable molecules from the trap, including those occupying $\ket{-}$. The pulse does not address any positive parity ground states so population in $\ket{+}$ remains shelved. From here, we apply an additional RF pulse to move the $\ket{+}$ population back to a detectable state before repumping and applying lambda cooling to image the molecules~\cite{hallas2022optical}. Additional details on the readout scheme can be found in the Supplemental Material~\cite{Supplemental}. The sequence is repeated for several Ramsey free precession times, $\tau$. The decay of the Ramsey fringe is fit to an exponential decoherence model to extract the coherence time $T_2^*$.

The $1/e$ lifetime of molecules in the $\tilde{X}(010)$ state is limited by blackbody radiation, radiative decay, and vacuum loss. It has been measured in our system to be $0.36_{-0.07}^{+0.11}$~s~\cite{hallas2022optical, vilas2023blackbody}. As this timescale is on the same order of the coherence times measured in this work, there is significant loss of coherent molecules during long Ramsey hold durations. In order to normalize out this background decay, we fix the length of the hold time in the bending mode to be independent of the Ramsey duration and apply ``pushout" pulses, so that we only detect molecules remaining in $\ket{+}$ following Ramsey. In this way, the average detectable molecular population following the Ramsey measurement is consistent across all the data.

\begin{figure}
    \centering
    \includegraphics{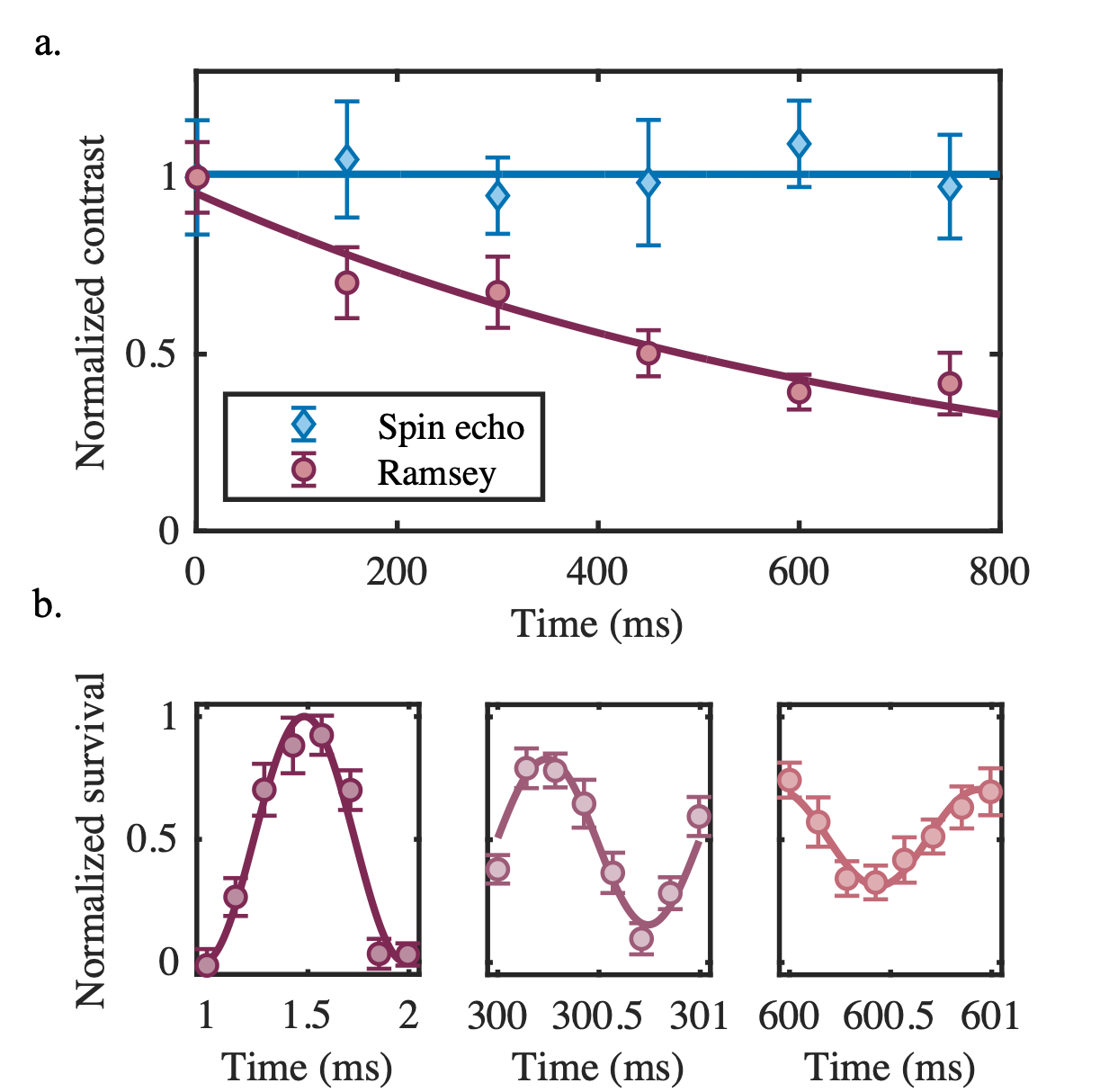}
    \caption{a. Coherence time of parity-doublet states in $N=1$. In the Ramsey measurement, two $\pi/2$ RF pulses are separated by a variable hold duration. We detune the RF frequency by 1~kHz, which allows us to measure over a period of 1~ms to extract the contrast of a full precession around the Bloch sphere. Decay of the contrast is fit to an exponential decoherence model to determine the coherence time, which we find to be $T_2^* = 0.8(2)~\text{s}$. 
    With spin echo, the coherence time is determined to be $>$$2.9$~s at the $95\%$ confidence level. The error reported comes from bootstrapped samples of the data.
    b. Three examples of the Ramsey measurement are shown below the coherence decay plot for free precessions $\tau$ of 1~ms, 300~ms, and 600~ms, normalized to the measured fringe contrast at 1~ms. Error bars represent $68\%$ confidence intervals.
    }
    \label{fig:2}
\end{figure}

We observe a coherence time of $T_2^* = 0.8(2)~\text{s}$ in the $N=1$ parity-doublet states. Figure \ref{fig:2} plots the decay of the Ramsey fringe contrast with free precession time as well as three examples of measured fringes. The observed dephasing is attributed to differential AC tensor Stark shifts from the trapping light, as discussed later in the text. As a consequence of the finite lifetime of the bending mode, observing coherence times longer than this timescale is challenging. Experimental improvements to the blackbody environment and vacuum lifetime could alleviate the loss rate to the bare radiative lifetime of the bending mode, $0.72_{-0.13}^{+0.25}$~s~\cite{hallas2022optical,zhang2025high}.

\begin{figure}
    \centering
    \includegraphics[width=\linewidth]{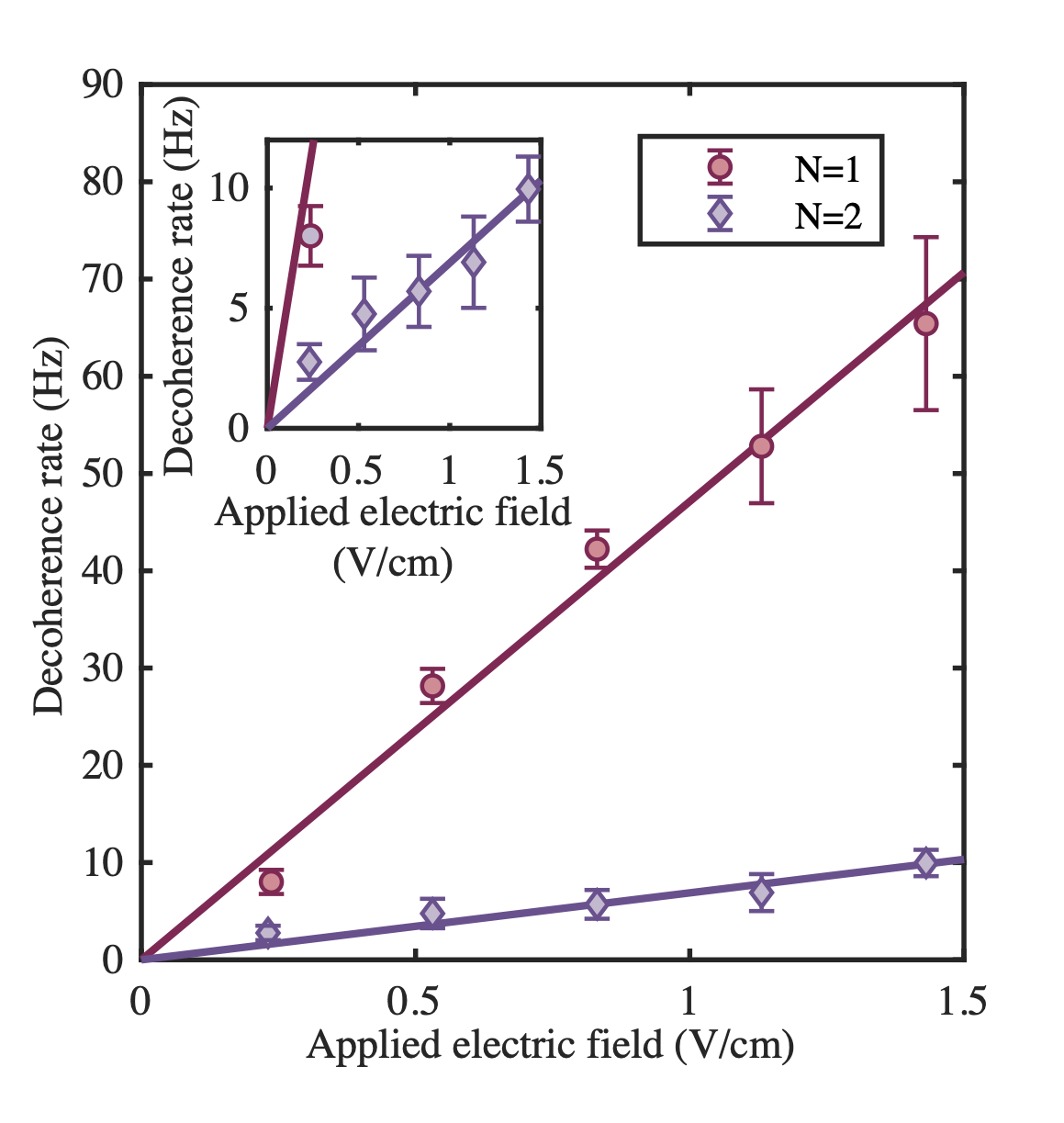}
    \caption{Decoherence rate dependence on electric field. We measure the Ramsey coherence time of parity-doublets in the $N=1$ and $N=2$ rotational states in several electric field environments. The decoherence rates shown are corrected for by the decoherence rate at zero applied field, which is limited by trap shifts. Due to the quadratic Stark shift, the decoherence dependence is fit to a line, as it reflects variation in the Stark shift. The ratio of the fitted slopes for $N=1$ and $N=2$ is 7(1), which is in agreement with our predicted value of 6.75 arising from the frequency splitting and dipole moment of the states. See the Supplemental Material for more details~\cite{Supplemental}. Error bars represent $68\%$ confidence intervals.}
    \label{fig:3}
\end{figure}

At low electric fields, the differential Stark sensitivity of the parity-doublet states is a quadratic shift~\cite{Supplemental}. To create a low electric field environment, we cancel static, ambient electric fields inside the vacuum chamber by applying bias voltages to in-vacuum coils and rod electrodes. Details on the cancellation process, which takes advantage of the Stark sensitivity of $\ket{1+}$ and $\ket{1-}$, can be found in the Supplemental Material~\cite{Supplemental}. The electric field environment can also fluctuate between experimental cycles, possibly arising from ions produced in the ablation plume of the cryogenic beam source adhering to insulating surfaces. The transient electric field is observed to be stabilized by applying $>$$100$~mW of ultraviolet light from UV LEDs at 275~nm and 365~nm during most of the experimental sequence, with the aim of inducing ion desorption from these surfaces. 

We map out the electric field sensitivity of the qubit states in $N=1$ and $N=2$ by applying a DC voltage across the in-vacuum MOT coils during the Ramsey measurement. Figure \ref{fig:3} plots the measured coherence time as a function of the applied electric field. Since the qubits experience a quadratic differential Stark shift at low electric fields, the decoherence rate varies linearly with applied field. Parity-doublet states in $N=1$ mix more strongly at lower electric fields compared to $N=2$ doublets as a consequence of their reduced energy spacing and larger dipole moment (see Supplemental Material~\cite{Supplemental}). Therefore, the $N=1$ doublets are expected to possess a greater sensitivity to fluctuations and offsets in the electric field environment, which is experimentally verified in Figure \ref{fig:3}. The slopes fit in Figure \ref{fig:3} provide information about the stability of the electric field environment in our system, which we deduce to be $\sim$$4$~mV/cm \cite{Supplemental}. Accounting for a differential quadratic shift of $\sim$$6.3$~kHz/(V/cm)$^2$ between parity-doublet states in $N=1$, we estimate overall absolute cancellation of the environmental electric field to within 20~mV/cm. These data show that the external electric field environment is not the dominant contribution to the observed dephasing in our system. Rather, we attribute the decoherence measured in Figure \ref{fig:2} to parity-dependent AC Stark shifts from the trapping light. 

\begin{figure}
    \centering
    \includegraphics[width=\linewidth]{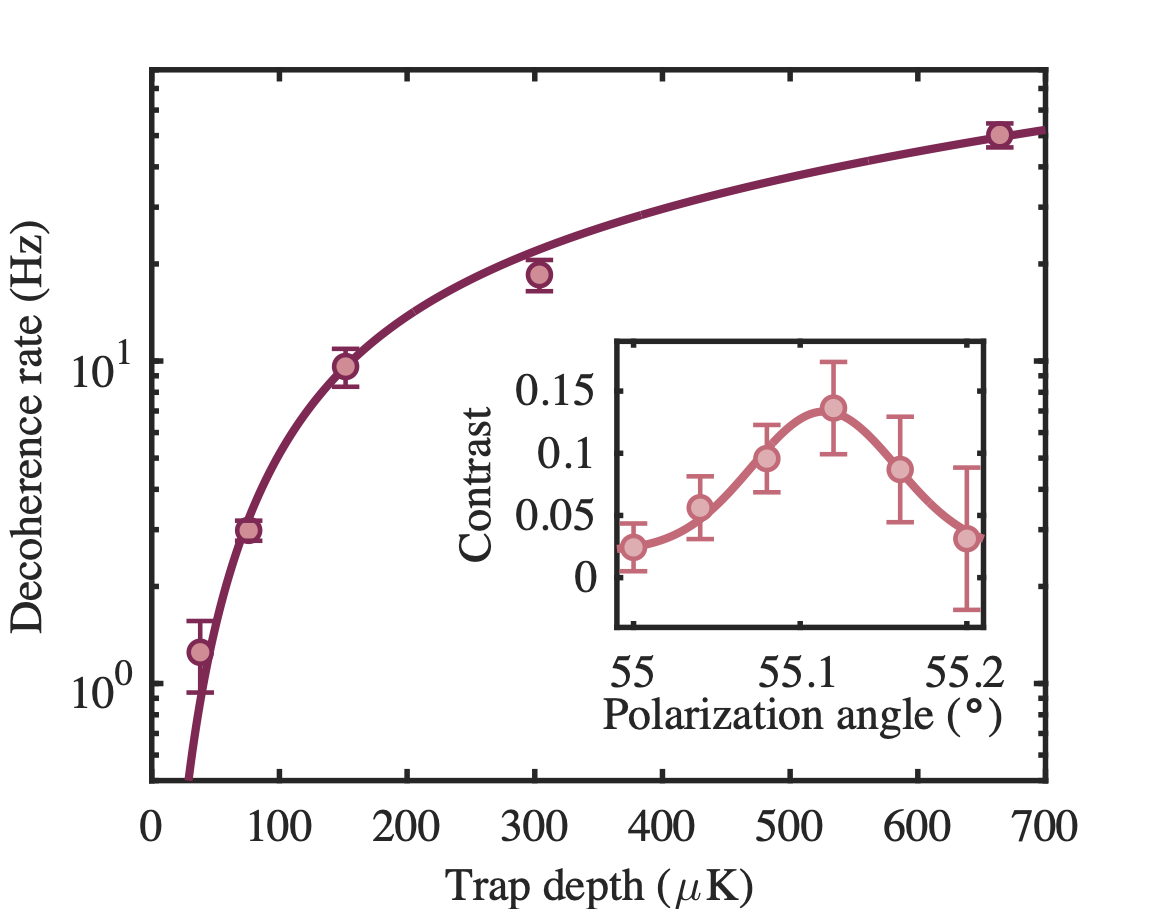}
    \caption{Decoherence rate dependence on trap depth. The decay in contrast for the Ramsey sequence is measured at several trap depths for $\ket{1+}$ and $\ket{1-}$. The temperature of the molecules during the Ramsey measurement is dependent on the trap depth as a result of adiabatic cooling and scales as $\propto\sqrt{U_0}$. The sensitivity of the decoherence rate to trap depth is numerically fit to a model which accounts for the spread of light intensities sampled by the molecules and the trap shift dependence (see Supplemental Material~\cite{Supplemental}). Inset: Magic angle scan. We measure the Ramsey contrast after $\tau = 0.45$~s as a function of trap polarization angle. The data shown here is for a trap depth of $U_0\sim40~\text{\textmu K}$. There is a calibration offset of the angle of up to 0.2\degree. Error bars represent $68\%$ confidence intervals.}
    \label{fig:4}
\end{figure}

While the optical trap is crucial to a long-lived molecular population, light shifts introduce a challenge to achieving long coherence times of parity doublets. The optical trap creates a confining force proportional to the polarizability of the molecule and the gradient of the light intensity. During the Ramsey measurement, molecules sample different intensities of the trap owing to their finite temperature. This spread of intensities, paired with a differential polarizability between parity-doublets, leads to decoherence. The differential polarizability is often quite large in molecules due to their anisotropic electronic wavefunctions, requiring careful strategies to minimize decoherence~\cite{neyenhuis2012anisotropic,gregory2017ac, Blackmore2018,seesselberg2018extending,Bause2020tuneout,guan2021magic,gregory2021robust,Burchesky2021Rotational, park2023extended,Gregory2024}. However, in $\ell$-type parity-doublet states, all angular momenta are identical, thereby eliminating AC Stark shifts associated with differential angular momenta coupling. The remaining, parity-dependent shifts arise predominantly from vibronic coupling in the electronic excited state \cite{Supplemental}. Notably, these shifts are over an order of magnitude weaker than in rotational qubits for diatomic molecules~\cite{Burchesky2021Rotational}.

To characterize the effect of trap shifts on the coherence of the parity-doublet states, we perform a Ramsey measurement at variable optical trap depths. The decoherence rate as a function of trap depth is plotted in Figure \ref{fig:4}. We note that the temperature of the molecules during the measurement is dependent on the trap depth, due to adiabatic cooling during the trap ramp. The differential light shift is mitigated by tuning the polarization angle of the optical trap with respect to the quantization axis set by the applied bias magnetic field~\cite{Burchesky2021Rotational}. At the ``magic" angle, the differential tensor polarizability between the parity doublet states is minimized. We ramp the bias field to 15-30~G preceding the measurement in order to resolve a well-defined magic condition. The magic angle condition is trap depth dependent. Thus, we optimize the angle for each coherence time measurement shown in Figure \ref{fig:4}. The inset in Figure \ref{fig:4} gives an example of the magic angle optimization for a trap depth $U_0\sim40~\text{\textmu K}$. The trap sensitivity of the decoherence rate is numerically fit to a model that includes a quadratic contribution from the magic angle condition in addition to a temperature-dependent effect reflecting the spread of light intensities traversed by the trapped molecules~\cite{Supplemental}.  

Transverse magnetic fields from the environment modify the optimal magic angle. Improved cancellation of transverse fields in the future should increase the measured coherence times. In addition, the sensitivity to transverse fields could be further reduced with a larger applied bias magnetic field. Improved cooling of the molecules would greatly improve the coherence time. Colder molecules sample a smaller variation in trap depth, and this would allow the trap to be ramped to lower intensities with less molecular loss.

We increase the observed coherence time by introducing a single spin-echo pulse. In the middle of the Ramsey hold we apply a $\pi$ pulse that acts to cancel out the dephasing from static inhomogeneities, observing no loss of contrast at $0.75$~s, as shown in Figure \ref{fig:2}. We measure a coherence time of $>$$2.9$~s at the $95\%$ confidence level. As the remaining population is limited by the finite lifetime of the bending mode, we do not measure coherence beyond this duration.

In conclusion, we demonstrate long parity-doublet coherence times in optically trapped CaOH molecules. We observe a quadratic Stark sensitivity between the parity-doublet qubit states, which is tunable by choice of rotational level. The differential magnetic field sensitivity of the states is minimal and is not limiting in this work. The observed dephasing of the parity-doublets is caused by a parity-dependent trap shift that is suppressed with a magic polarization angle. Residual ambient transverse magnetic fields limit the achievable coherence in our current system. We expect longer coherence times to be feasible with a larger bias magnetic field and improved transverse magnetic field stabilization. Tuning the trapping light to a magic wavelength, such as near a weak vibrational transition, could further cancel the parity-dependent light shift and improve the coherence time of the doublets~\cite{Bause2020tuneout, guan2021magic, Gregory2024}.


The long parity-doublet coherence times in this work can be naturally extended from the large optical trap used here to optical tweezer arrays, a platform already demonstrated with CaOH molecules~\cite{VilasOpticalTweezer2024}. The single particle control of a tweezer array enables engineered dipolar spin-exchange interactions between single molecules in parity-doublet states as a next step. Entanglement between polyatomic molecules on a time scale much shorter than qubit decoherence is a key requisite for a number of applications, extending the work with diatomic molecules to a new complexity regime~\cite{holland2023demand,bao2023dipolar,picard2025entanglement,ruttley2025long}. Parity-doublet molecules pinned in an optical lattice make a natural analog system for novel quantum simulations of integer-spin models~\cite{Wall2013}. The long coherence times observed in this work make these studies particularly promising. Furthermore, long bare coherence times, for example in heavy molecules such as SrOH and RaOH~\cite{kozyryev2017PolyEDM,Norrgard2019,kozyryev2021enhanced,zhang2023relativistic,Arrowsmith-Kron2024opportunities}, offer promise as a powerful tool for precision searches for BSM physics. Ultimately, the utility of parity-doublet states in linear triatomic molecules is limited by the finite effective lifetime of the bending mode~\cite{hallas2022optical, vilas2023blackbody}. Blackbody losses contributing to this lifetime can be suppressed in a cryogenic experimental environment~\cite{zhang2025high}.  

Asymmetric top molecules (ATMs) present another, even richer, complexity frontier. Parity-doublet states are generic to ATMs in their vibrational {\sl{ground}} state and offer $\gg10$~s lifetimes. Recent work has identified several ATMs as promising candidates for laser cooling~\cite{kozyryev2016MOR, Augenbraun2020ATM, Seanthesis,frenett2024vibrational}. Their large intrinsic electric dipole moments would enable entangling gates on timescales far shorter than single particle coherence times, making parity-doublet states in ATMs another promising direction for quantum science and precision searches for BSM physics~\cite{kozyryev2016MOR,Hutzler2020, Augenbraun2020ATM}.

We thank Avikar Periwal and Abdullah Nasir for valuable discussions. This material is based upon work supported by the Air Force Office of Scientific Research (Award Numbers FA9550-22-1-0228 and FA2386-24-1-4070), the National Science Foundation (PHY-2409404 and NSF Physics Frontier Center PHY-2317134), and the U.S. Army Research Office (W911NF1910283). Support is also acknowledged from the U.S. Department of Energy, Office of Science, National Quantum Information Science Research Centers, Quantum Systems Accelerator. 

P.R., C.H., J.T., G.L., and N.B.V. performed the experiment and analyzed the data. J.M.D. directed the study. All authors discussed and developed understanding of the results, and contributed to the manuscript.

\bibliography{CaOHReferences}

\title{Methods}


\maketitle
\setcounter{figure}{0}
\renewcommand{\thefigure}{S\arabic{figure}}
\newpage

\section{Experimental details}

The experimental apparatus has been described in previous publications~\cite{vilas2022magneto}. In this work, we start by loading CaOH molecules from a blue-detuned magneto-optical trap into an optical dipole trap (ODT) using single-frequency gray molasses cooling~\cite{hallas2022optical, Hallas2024, vilas2025collisions}. The trap depth of the ODT is $\sim$$640~\text{\textmu K}$. After loading the ODT, molecules are imaged with $10~\text{ms}$ of single-frequency light to provide a measurement for the number of loaded molecules~\cite{hallas2022optical}. An additional pulse of single-frequency cooling light optimized for cooling, not imaging, is then used to cool the molecules to $\sim$$60~\text{\textmu K}$. The cooled molecules are optically pumped into the bending mode $\tilde{X}(010)$, which contains parity-doublet states, via the transition $\tilde{X}(000)\rightarrow\tilde{A}(010)\kappa^2\Sigma^{(-)}$~\cite{anderegg2023quantum}. Subsequently, we apply a sequence of frequency-modulated microwave and optical pulses to prepare the population in a single hyperfine state, $|J=1/2, F=0, m=0, p=-\rangle$, as described in previous work~\cite{anderegg2023quantum}. A schematic of the experimental sequence to measure coherence in the $N=1$ parity doublets is given in Figure~\ref{fig:S.1}.

\subsection{State preparation}

Once in a single hyperfine state, molecules are driven to the $\ket{1+}$ state with an RF pulse. Since the transition $|N=1, J=1/2, F=0, m=0, p=-\rangle \rightarrow |N=1, J=3/2, F=2, m=2, p=+\rangle$ is only allowed due to mixing from the trapping light, it is relatively weak and dependent on characteristics of the trapping light, e.g., the polarization angle. In order for the state preparation to be robust across the magic polarization angles used in this work, we drive the transition using adiabatic rapid passage (ARP). By sweeping the magnetic field from $0.9~\text{G}$ to $3~\text{G}$ and applying RF over 10~ms, we transfer $>$95\% of the population to the $\ket{1+}$ state. Any population remaining in $\ket{N=1, J=1/2, F=0, p=-}$ is depleted with a ``pushout'' pulse of resonant light and repumping light. We again note here that positive parity ground states are not addressed by any of the vibrational repumping lasers. Following ARP to $\ket{1+}$, we prepare for the Ramsey sequence by ramping the trap to a desired trap depth while ramping the magnetic field to $\sim30~\text{G}$. At the lowest trap depth explored in this work, $\sim$$40~\text{\textmu K}$, the reduction in trap depth results in a $\sim$$50\%$ loss of molecules from the trap.

For measurements of the $N=2$ parity-doublet coherence times, we follow identical steps as described above. After the trap and magnetic field ramp, we apply a MW $\pi$ pulse that drives $|N=1, J=3/2^+,F=2,m=2\rangle\rightarrow |N=2,J=5/2^-,F=3,m=3\rangle$, preparing the $\ket{2-}$ state.

\subsection{Vacuum cancel}
Preceding the Ramsey sequence, we apply a variable hold duration, which we refer to as a ``vacuum cancel''. The intention of this hold is to mitigate the disparate effects of blackbody radiation and radiative decay that disproportionately affect longer Ramsey hold times. In previous work, we have measured~\cite{hallas2022optical} and characterized~\cite{vilas2023blackbody} the effects of radiative decay, blackbody thermalization, and vacuum loss on the lifetime of the CaOH bending mode in our system. The effective lifetime of the state is $0.36_{-0.07}^{+0.11}$~s, with the most significant contributions from blackbody radiation ($0.90_{-0.16}^{+0.4}$~s) and radiative decay of the state ($0.72_{-0.13}^{+0.25}$~s)~\cite{hallas2022optical,vilas2023blackbody}. Therefore, we observe significant loss of molecules in the bending mode for hold times on the order of 100~ms. 

\begin{figure*}
    \centering
    \includegraphics[width=\textwidth]{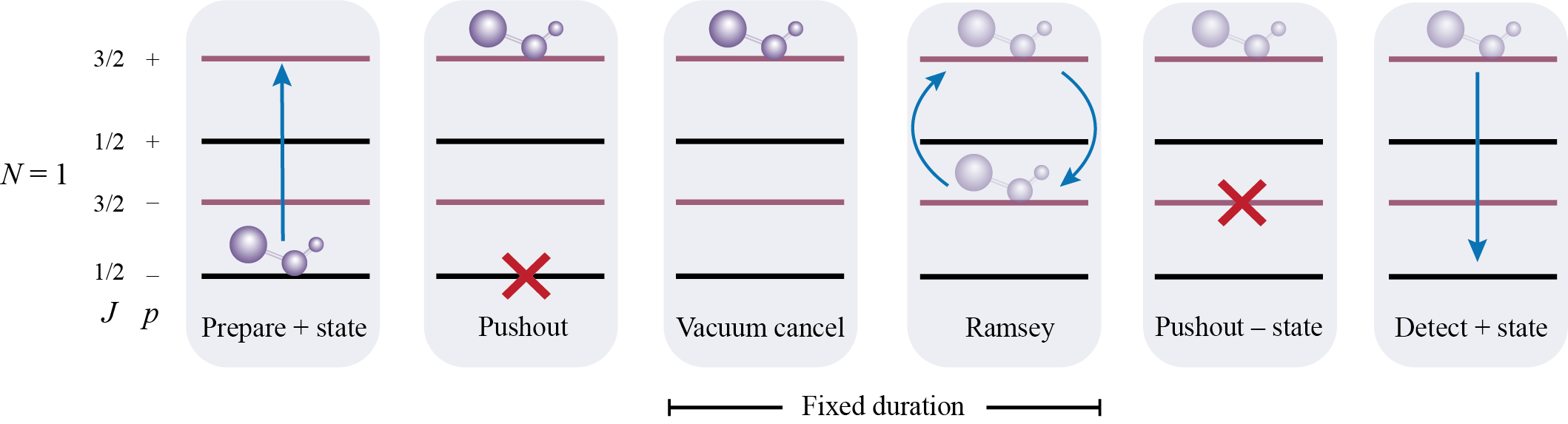}
    \caption{Experimental sequence used in this work to determine the coherence time for $N=1$ parity-doublet states, as read left to right. Pushout pulses use resonant mainline light to deplete all molecules in negative parity ground states.}
    \label{fig:S.1}
\end{figure*}

Without correcting for this effect, the measured decay in Ramsey contrast would reflect both a dephasing of the parity-doublet states in addition to a loss of molecular population. The timescales of these contributions are on the same order of $\sim$100~ms. To correct for the loss in our system, we implement a variable vacuum cancel hold preceding the Ramsey measurement so that the total hold duration in the bending mode is fixed and independent of the free precession time (see Figure~\ref{fig:S.1}).

\subsection{Readout}
Following the Ramsey sequence, we ramp the magnetic field back to 3~G and the trap depth up to $U\sim640~\text{\textmu K}$. We then apply a pushout pulse to deplete molecular population in $\ket{1-}$. To detect population remaining in $\ket{1+}$, we transfer molecules back to $|N=1, J=1/2^-,F=0\rangle$ with the RF ARP sweep in reverse. The molecules are then imaged with 40~ms of $\Lambda$-cooling light~\cite{hallas2022optical}. 

For the $N=2$ parity-doublet states, we ultimately detect population remaining in $\ket{2+}$. While neither $\ket{2+}$ nor $\ket{2-}$ is addressed in our photon cycling scheme, population from these states can decay or be driven by blackbody radiation to detectable states. For this reason, we start by depleting population in the $\ket{2-}$ state by driving the MW $\pi$ pulse to $N=1$ and RF ARP in reverse, before applying a pushout pulse. Subsequently, we drive $\ket{2+}$ population to $|N=1, J=3/2^-\rangle$ with a MW $\pi$ pulse, before imaging.

\section{Electric field cancellation}
At low electric fields, parity-doublet states begin to mix, which leads to a quadratic Stark shift. To demonstrate long coherence times in CaOH, it is crucial to suppress the Stark sensitivity by engineering a near-zero electric field environment. In our system, stray electric fields can arise from surface charges on the insulating surfaces inside the vacuum chamber, including an aspheric lens and aluminum nitride MOT coil boards. We can cancel static electric fields produced by these charges in all directions by applying compensation voltages to the in-vacuum MOT coils and four in-vacuum electrode rods. To determine what applied voltages are necessary to cancel the electric field environment, we utilize the Stark sensitivity of the parity doublets. The Stark shift corresponds to an increase in the transition frequency between the parity-doublet states. Therefore, canceling the environmental electric fields is analogous to minimizing the parity-doublet transition frequency.

\begin{figure*}
    \centering
    \includegraphics[width=\textwidth]{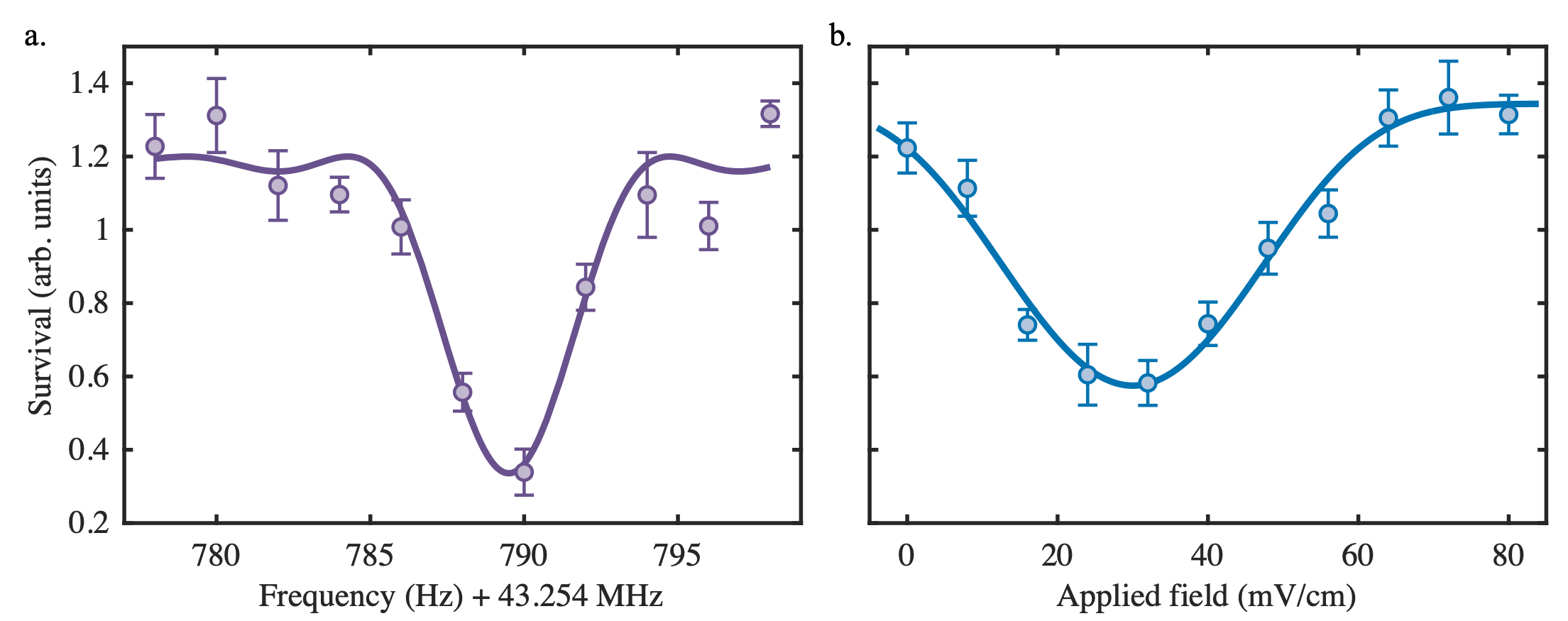}
    \caption{a. Rabi spectroscopy of the parity-doublet transition $\ket{1+}\rightarrow\ket{1-}$ with a low energy drive ($\Omega = 5.5$~Hz). The transition is sensitive to fluctuations and offsets in the electric field environment. We are able to cancel static, ambient fields in the chamber by minimizing the transition frequency, using the Hz-level Rabi spectroscopy shown here.  
    b. Minimization of the electric field environment along the $y$-axis in our system. The RF transition frequency is chosen to be 2~Hz below the resonance frequency in the found from the Rabi spectroscopy. In this way, the electric field minimum, which occurs at the lowest transition frequency, corresponds to the lowest value in the plot. Here we find that a $30~\text{mV/cm}$ compensation field best cancels the field along $y$.}
    \label{fig:S.2}
\end{figure*}

We cancel the stray electric fields by performing high-resolution, Hz-level spectroscopy on the trapped molecules with a long, low energy $\pi$ pulse on the $N=1$ parity-doublet transition. Maximum contrast is achieved at the resonant frequency, shown in Figure \ref{fig:S.2}a. The spectroscopy curve could, in principal, be iteratively taken as a function of applied voltage. For efficiency, we instead drive the $\pi$ pulse at a frequency lower than the resonance measured and scan the compensation voltage. In this way, the minimum electric field environment will correspond to the largest contrast, as shown in Figure \ref{fig:S.2}b. We perform this procedure iteratively and find minimal coupling between the three axes of our system, set by the MOT coils and 2 orthogonal sets of electrode rods. To minimize the overall field environment, the compensation voltages across each pair of rod electrodes are identical and opposite. 

We find the static electric field environment of our system approximately stable on the time scale of a few hours to a day. While day-to-day the compensation fields can change up to $\sim$20~mV/cm, we are able to consistently cancel fields to the same resonance transition frequency, within 1~Hz. Ultraviolet LEDs of $>$100~mW at 275~nm and 365~nm help maintain the stability of the field environment by inducing desorption of charges inside the vacuum chamber. We find that more UV power only improves the stability of the system. The slope in Figure 3 for the $N=1$ parity doublets is fit to be 46.81~Hz/V/cm. This value implies fluctuations in the electric field environment of $\sim$4~mV/cm in our system. With this information, we deduce that the electric field at the molecules is $<$20~mV/cm. 

\section{Rotational Stark sensitivity}
The electric field response of the stretched parity-doublet states explored in this work can be simplified as a two-level system coupled by a field $\mathcal{E}$ and with an energy splitting $\Delta$. The resulting eigenvalues of the system are given by 
$$
    E_{\pm} = \pm \frac{\hbar\Delta}{2}\sqrt{1+4\left(\frac{\mu \mathcal{E}}{\hbar\Delta}\right)^2},
$$ 
where $\mu$ is the transition dipole moment between the parity-doublet states. At low electric fields ($\mu\mathcal{E}\ll\hbar\Delta$), these energies can be expressed as $E_\pm \approx \pm \hbar\Delta/2 \pm (\mu \mathcal{E})^2/\hbar\Delta$. The slope fit in Figure 3 is related to the derivative of this energy with respect to the electric field, which is simply $2 |\mu|^2/\hbar\Delta$. Intuitively, the electric field sensitivity is dependent on the strength of the interaction between the states and their energy splitting.

Both $\mu$ and $\Delta$ are dependent on the rotational state of the molecule. In the fully stretched states in the bending mode of a linear polyatomic molecule~\cite{Hirota1985}, 
$$
    \mu = \mu_{\text{el}} \frac{|\ell|}{N+1}.
$$
The dipole moment is therefore $\mu_{\text{el}}/2$ for the $\ket{\pm1}$ doublets and $\mu_{\text{el}}/3$ for the $\ket{\pm 2}$ pair. The energy splitting of the parity-doublet states in CaOH is dependent on the $\ell$-doubling parameter $q$ and the rotational quantum number, and is given by $qN(N+1)$ \cite{coxon1994laser,li1995high}.

With this approach, we predict the ratio of the electric field sensitivity for fully-stretched parity doublets in $N=1$ to $N=2$ to be 27/4 = 6.75, which is in agreement with the experimentally determined value of 7(1), visualized in Figure 3.

\section{Trap shift sensitivity}
Here we derive the scaling of decoherence with trap intensity, which is fit to the data in Figure 4 of the main text. Based on solutions of the molecular Hamiltonian, which includes AC Stark shifts from the ODT light, we find that there is a small differential light shift between parity-doublet states that varies quadratically with the trap intensity. This dependence is similar to the differential trap shifts observed in CaF, where the linear polarizability and hyperpolarizability have competing effects~\cite{Burchesky2021Rotational}. We denote the differential light shift as $\Delta f$. 

Trapped molecules sample a distribution of ODT light intensities as a result of their finite temperature. Due to the change in temperature from adiabatic cooling during the trap ramp preceding the measurement, the width of the distribution of $\Delta f$ is dependent on the final trap depth (the initial trap depth is fixed for all data). The magic angle condition also depends on the final trap depth due to a combination of the quadratic differential shift and the change in temperature.
In order to capture the dependence of these effects on trap depth, we estimate the decoherence rate as the standard deviation of $\Delta f$, i.e., $1/T_2^* \propto \sigma(\Delta f)$.
We parameterize $\Delta f$ as
$$
    \Delta f(I) = a\left(2 I_{\text{magic}}I-I^2\right),
$$
where $a$ is a constant proportional to the polarizability of the molecule, $I$ is the intensity of the trap light, and $I_{\text{magic}}$ is defined as the intensity at the zero-slope point of the differential shift, i.e., $(d/dI)\Delta f(I_\text{magic}) = 0$. $I_{\text{magic}}$ is a function of $I_0$, the maximum intensity of the trapping light, which is related to the trap depth as
$$
    U_0 = -\frac{\alpha}{2\epsilon_0c} I_0,
$$
where $\alpha$ is the polarizability, $\epsilon_0$ is the electric constant, and $c$ is the speed of light. We can express the width of the $\Delta f$ sampled by the molecules as 
$$
    \sigma(\Delta f) = \sqrt{\mathbb{E}[\Delta f^2] - \mathbb{E}[\Delta f]^2},
$$
where $\mathbb{E}[\cdot]$ is the expectation value operator. To evaluate $\sigma(\Delta f)$, we use that the probability density function $p$ of energies is given by the  Boltzmann distribution~\cite{Kuhr2005analysis,Tuchendler2008energy}. Written in terms of the trap intensity, this function is
$$
    p(I) = \frac{2\eta^3}{\beta I_0}\left(1-\frac{I}{I_0}\right)^2\exp\left(-2\eta\left(1-\frac{I}{I_0}\right)\right)
$$ 
for $I > 0$, where $\beta = \alpha/2\epsilon_0 c$ and $\eta = U_0/k_BT$. The data is fit to the model with free parameter $a$.

\end{document}